%% file: main.tex
\documentclass[10pt,a4paper,twocolumn]{article}

\usepackage[utf8]{inputenc}
\usepackage[T1]{fontenc}
\usepackage{mathptmx}  
\usepackage{amsmath,amssymb,amsfonts}
\usepackage{graphicx}
\usepackage{booktabs}
\usepackage{multirow}
\usepackage{hyperref}
\usepackage{cleveref}
\usepackage{xcolor}
\usepackage{soul}
\usepackage{microtype}
\usepackage{natbib}
\usepackage{algorithm}
\usepackage{algorithmic}
\usepackage{enumitem}
\usepackage{subcaption}
\usepackage{tabularx}
\usepackage{adjustbox}
\usepackage{fancyvrb}
\usepackage{caption}
\usepackage{float}
\usepackage{placeins}
\usepackage[top=1in,bottom=1in,left=0.75in,right=0.75in]{geometry}

\hypersetup{
    colorlinks=true,
    linkcolor=blue,
    citecolor=blue,
    urlcolor=blue
}

\title{Trust, Lies, and Long Memories: Emergent Social Dynamics and Reputation in Multi-Round Avalon with LLM Agents}

\author{
    Suveen Ellawela \\
    \texttt{suveen.ellawela@u.nus.edu} \\
    National University of Singapore
}

\date{}

\begin{document}

\onecolumn

\begin{center}
{\LARGE\bfseries Trust, Lies, and Long Memories: Emergent Social Dynamics and Reputation in Multi-Round Avalon with LLM Agents\par}
\vspace{0.5cm}
{\large Suveen Ellawela\\
\texttt{suveen.ellawela@u.nus.edu}\\
National University of Singapore\\
\url{https://github.com/SuveenE/multi-round-avalon-agents}\par}
\end{center}
\vspace{0.3cm}

\begin{figure}[H]
    \centering
    \includegraphics[width=\textwidth]{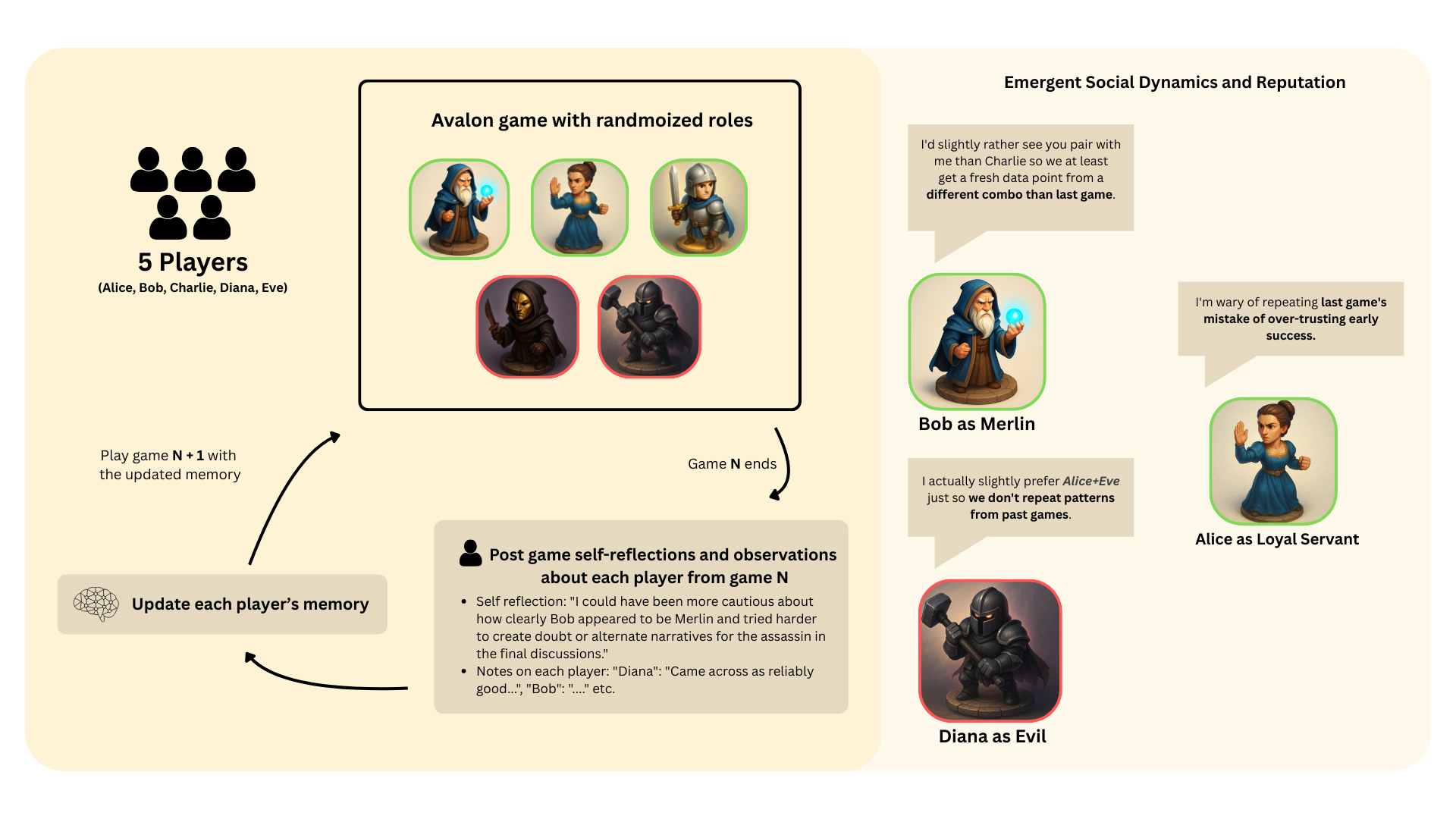}
    \caption{Overview of our experimental setup showing emergent social dynamics and reputation. Five LLM agents (Alice, Bob, Charlie, Diana, Eve) play repeated Avalon games with randomized roles. After each game, agents generate self-reflections and observations about other players, which persist into subsequent games as memory. This cross-game memory enables reputation formation, with agents referencing past behavior in their strategic reasoning (example quotes shown on right).}
    \label{fig:overview}
\end{figure}

\begin{center}
\begin{minipage}{0.95\textwidth}
\textbf{Abstract.}
We study emergent social dynamics in LLM agents playing The Resistance: Avalon, a hidden-role deception game. Unlike prior work on single-game performance, our agents play repeated games while retaining memory of previous interactions (who played which roles and how they behaved), enabling us to study how social dynamics evolve. Across 188 games, two key phenomena emerge. First, \textbf{reputation dynamics emerge organically} when agents retain cross-game memory: agents reference past behavior in statements like ``I'm wary of repeating last game's mistake of over-trusting early success.'' These reputations are role-conditional: the same agent is described as ``straightforward'' when playing good but ``subtle'' when playing evil, and high-reputation players receive 46\% more team inclusions. Second, \textbf{higher reasoning effort supports more strategic deception}: evil players more often pass early missions to build trust before sabotaging later ones (75\% in high-effort games vs 36\% in low-effort games). Together, these findings show that repeated interaction with memory gives rise to measurable reputation and deception dynamics among LLM agents.

\vspace{0.3cm}
\end{minipage}
\end{center}

\vspace{0.5cm}

\twocolumn

\input{sections/introduction}
\input{sections/related_work}
\input{sections/game_environment}
\input{sections/methods}
\input{sections/results}
\input{sections/discussion}
\input{sections/conclusion}

\bibliographystyle{plainnat}
\bibliography{references}

\appendix
\input{sections/appendix}

\end{document}

%% file: sections/introduction.tex
\section{Introduction}
\label{sec:introduction}

Social deduction games provide a compelling testbed for studying emergent behavior in AI systems. Unlike chess or Go, where optimal play can be computed, games like The Resistance: Avalon \citep{wiki:resistance} require \textit{social reasoning}: inferring hidden information from behavior, building trust, detecting deception, and coordinating without explicit communication channels. These capabilities are central to real-world multi-agent scenarios, from negotiation to collaborative problem-solving.

Recent advances in large language models (LLMs) have enabled agents that can engage in extended natural language dialogue, maintain context across interactions, and reason about others' mental states \citep{kosinski2024tom}. This raises fundamental questions about multi-agent multi-round dynamics: Do LLMs develop stable social models when they interact repeatedly with the same agents across multiple games? Are these models role-conditional, where agents distinguish how the same individual behaves as deceiver versus cooperator?

We address these questions using The Resistance: Avalon, a hidden-role game where a ``good'' team attempts to complete missions while ``evil'' players sabotage secretly. The game includes an \textit{assassination} mechanic: if the good team wins, evil can still claim victory by identifying ``Merlin,'' a good player who knows evil's identities but must hide this knowledge.

Our contributions are as follows:
\begin{itemize}[nosep]
    \item We show that LLM agents develop stable, role-conditional reputations when given cross-game memory: the same player is described as ``subtle'' when evil but ``straightforward'' when good. Agents actively reference previous games and their learnings when making decisions.
    \item We find that high-reputation players receive 45.6\% more team inclusions, showing that reputation has downstream strategic consequences for coalition formation.
    \item We find that higher reasoning enables more sophisticated deception: evil players more frequently pass early missions to build trust before sabotaging, a strategy known as ``sleeper agent'' among human players (75\% at medium/high reasoning vs 36\% at low reasoning).
\end{itemize}

%% file: sections/related_work.tex
\section{Related Work}
\label{sec:related}

\subsection{LLMs in Social Deduction and Board Games}
\label{sec:related:avalon}

Most directly related to our work is \textbf{AvalonBench} \citep{light2023avalonbench}, which introduced a benchmark for evaluating LLM agents in The Resistance: Avalon. AvalonBench focused on single-game performance with rule-based opponents; our work extends this in three key directions, motivated by bringing the game closer to how humans actually play it through multi-agent multi-round dynamics. When humans play repeatedly with the same group, they develop mental models of how each player behaves differently depending on their team assignment, noticing that someone might be ``aggressive when evil'' but ``cautious when good.'' First, we introduce cross-game memory to study how such reputation and trust dynamics develop over repeated interactions between the same LLM agents. Second, we systematically vary reasoning depth to isolate its effect on strategies used during game play. Third, we conduct detailed qualitative analysis of the natural language strategies that emerge, cataloging role-conditional reputation patterns and behavioral ``tell'' taxonomies that agents use to identify hidden roles.

\subsection{LLMs in Strategic Multi-Agent Settings}
\label{sec:related:strategic}

The broader literature on LLMs in strategic settings has examined negotiation, where \citet{lewis2017negotiation} trained end-to-end dialogue agents for deal-making, and diplomacy, where Meta's Cicero \citep{meta2022cicero} achieved human-level play by combining language models with strategic planning. \citet{park2023generative} demonstrated that LLM agents in simulated social environments develop emergent behaviors like information diffusion and relationship formation. The AgentBench framework \citep{liu2023agentbench} provides comprehensive evaluation of LLM agents across code generation, web browsing, and other single-agent environments, establishing that frontier models substantially outperform smaller models on agentic tasks. However, multi-agent social deduction, where agents must reason about hidden information, form coalitions, and detect deception through behavioral analysis, remains underexplored. Our work addresses this gap by studying repeated games with persistent memory, enabling investigation of how agents build, maintain, and exploit social models over time.

\subsection{Deception and Theory of Mind}
\label{sec:related:deception}

Detection of deceptive behavior in AI systems is an active area of AI safety research. \citet{hubinger2024sleeper} demonstrated that deceptive behaviors can persist through safety training, raising concerns about detecting misaligned AI systems. Avalon provides a complementary perspective: a controlled setting where ground truth (role assignments) is known, enabling precise measurement of detection accuracy and systematic analysis of failure modes. On the cognitive side, \citet{kosinski2023tom} found evidence that theory of mind capabilities may emerge in large language models, with GPT-4 passing classic false-belief tasks. Our work extends this to dynamic social contexts where mental models must be continuously updated based on behavioral evidence across extended multi-party interactions, a substantially more demanding test of social cognition than static vignette-based assessments.

%% file: sections/game_environment.tex
\section{Game Environment: The Resistance: Avalon}
\label{sec:game}

\begin{figure*}[t]
    \centering
    \includegraphics[width=0.85\textwidth]{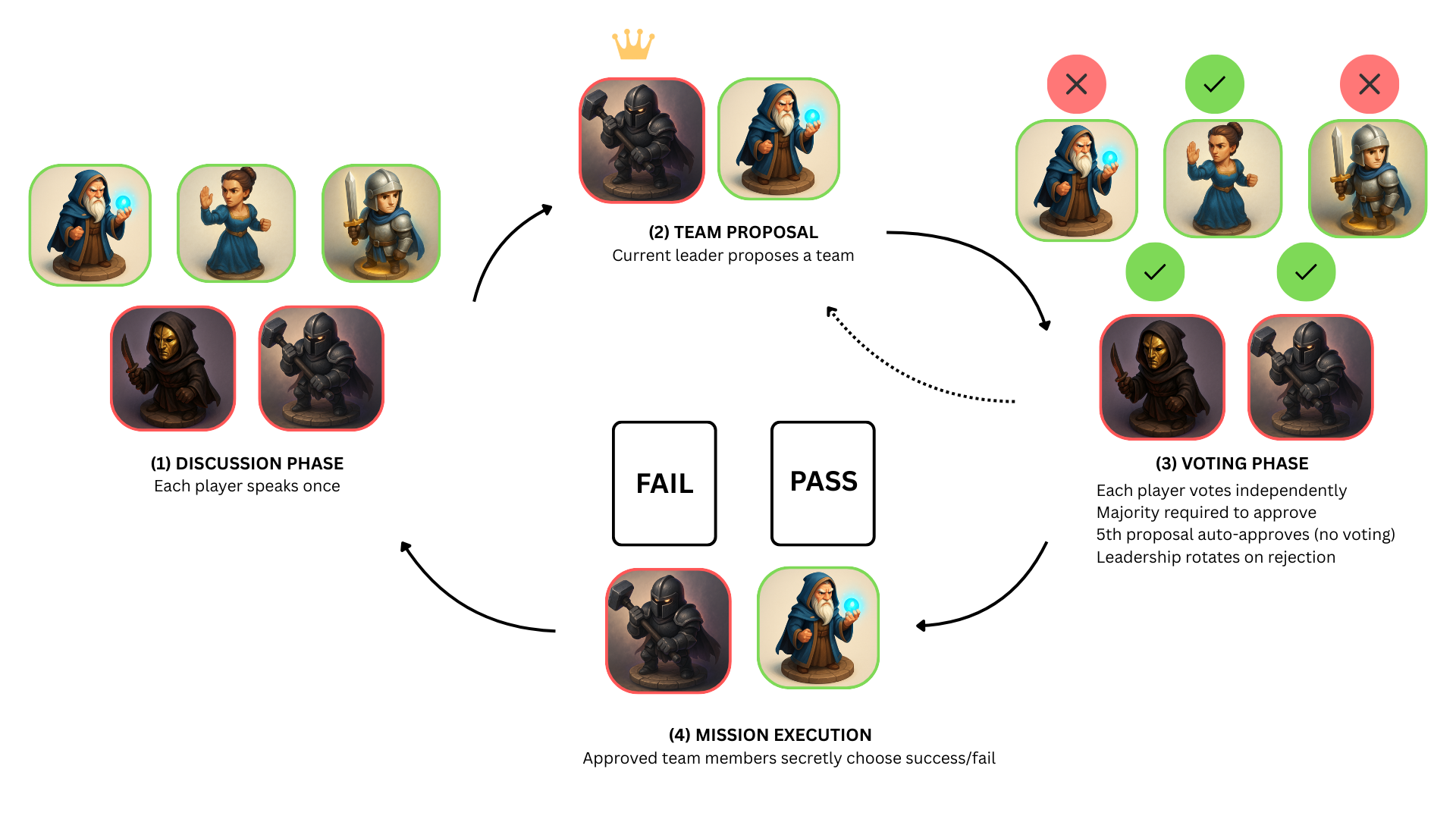}
    \caption{Game flow in The Resistance: Avalon. Each mission round consists of four phases: (1) Discussion, where players share observations and suspicions; (2) Team Proposal, where the current leader selects team members; (3) Voting, where all players approve or reject the proposal; and (4) Mission Execution, where approved team members secretly choose success or fail. The cycle repeats until one team wins three missions.}
    \label{fig:gameflow}
\end{figure*}

This section provides a detailed description of The Resistance: Avalon, the hidden-role social deduction game used in our experiments.

\subsection{Overview and Objectives}
\label{sec:game:overview}

The Resistance: Avalon is a game of hidden loyalties for 5--10 players. Players are secretly divided into two teams: the \textbf{Good Team} (Loyal Servants of Arthur), who form the majority and aim to successfully complete three of five missions, and the \textbf{Evil Team} (Minions of Mordred), a minority who know each other's identities and aim to either sabotage three missions or identify and assassinate Merlin. The game captures a fundamental tension: good players must coordinate without certain knowledge of who to trust, while evil players must deceive while appearing trustworthy.


\subsection{Special Roles}
\label{sec:game:roles}

Beyond basic good/evil alignment, Avalon includes special roles with unique information or abilities:

\begin{table}[h]
\centering
\caption{Special roles in The Resistance: Avalon.}
\label{tab:roles}
\footnotesize
\begin{tabular}{@{}llp{3cm}@{}}
\toprule
\textbf{Role} & \textbf{Team} & \textbf{Ability} \\
\midrule
Merlin & Good & Knows evil; must guide subtly \\
Percival & Good & Sees Merlin \& Morgana \\
Assassin & Evil & Can assassinate Merlin \\
Morgana & Evil & Appears as Merlin \\
Mordred & Evil & Hidden from Merlin \\
Oberon & Evil & Isolated from evil team \\
\bottomrule
\end{tabular}
\end{table}

The interplay of these roles creates rich strategic complexity. Merlin must guide without revealing; Percival must protect without certainty; evil must deceive while coordinating.

\subsection{Team Composition}
\label{sec:game:composition}

The ratio of good to evil players varies with player count. In the 5-player variant used in most of our experiments, the Good Team consists of Merlin and 2 Loyal Servants, while the Evil Team consists of the Assassin and 1 generic Minion. Role configurations for other player counts (6--10 players), which introduce additional special roles like Percival, Morgana, Mordred, and Oberon, are provided in \Cref{sec:appendix:composition}.

\subsection{Game Flow}
\label{sec:game:flow}

A game of Avalon proceeds through the following phases:

\paragraph{Phase 1: Role Assignment and Night Phase}
At game start, each player is secretly assigned a role. During the ``night phase,'' players receive their private information: evil players (except Oberon) learn each other's identities, Merlin learns which players are evil (except Mordred), and Percival learns which players appear as Merlin (the true Merlin plus Morgana, without knowing which is which). This information asymmetry is the foundation of all strategic interaction.

\paragraph{Phase 2: Mission Rounds (Repeated up to 5 times)}
The game consists of up to 5 mission rounds. Each round proceeds through four steps:
\begin{enumerate}[nosep]
    \item \textbf{Team Proposal:} The current Leader proposes a team of specified size for the mission and explains their reasoning.
    \item \textbf{Discussion:} All players debate the proposed team, arguing for or against inclusion of specific individuals. This is the primary opportunity for information exchange and deception.
    \item \textbf{Team Vote:} All players simultaneously vote to Approve or Reject; if a majority approves, the team proceeds to the mission, while rejection passes leadership clockwise. After 4 consecutive rejections, the 5th proposal auto-approves.
    \item \textbf{Mission Execution:} Approved team members secretly choose Success or Fail. Good players must choose Success, while evil players may choose either strategically. Any Fail card fails the mission, except Mission 4 in 7+ player games which requires 2 Fails.
\end{enumerate}

In a 5-player game, missions require teams of 2, 3, 2, 3, and 3 players respectively. Full mission team sizes for all player counts are provided in \Cref{sec:appendix:composition}.

\paragraph{Phase 3: Victory Determination}
The game ends when either team achieves victory. Good wins by successfully completing 3 of 5 missions. Evil wins either by failing 3 missions or through assassination: if good wins by missions, the Assassin gets one chance to identify Merlin, and a correct guess gives evil the victory instead.

%% file: sections/methods.tex
\section{Methods}
\label{sec:methods}

Our experimental framework enables LLM agents to play repeated Avalon games while retaining memory of past interactions. This section describes the agent architecture, memory system, reasoning manipulation, and data collection procedures.

\subsection{Agent Architecture}
\label{sec:methods:architecture}

Following the ReAct paradigm \citep{yao2023react}, each agent is instantiated as a prompted LLM (OpenAI GPT-5.1 \citep{openai2025gpt51}) that interleaves reasoning with action. Each agent receives:
\begin{itemize}[nosep]
    \item \textbf{Role knowledge:} Private information about their role and, for some roles, others' identities
    \item \textbf{Game state:} Public mission history, vote records, and discussion transcripts
    \item \textbf{Memory (tournament mode):} Reflections from previous games including observations about other players
\end{itemize}
Agents generate natural language discussion contributions, vote on team proposals, and (if on approved mission teams) choose to succeed or fail the mission.

\subsection{Memory System}
\label{sec:methods:memory}

In tournament mode, agents retain memory across games through a structured reflection system. At the end of each game, all roles are revealed to all agents (mirroring standard human play). Agents then generate a post-game reflection containing a self-assessment of their own performance and observations about each other player's behavior; these reflections may therefore include explicit role identities (e.g., ``Bob was evil this game''). At the start of subsequent games, agents receive their last 3 self-assessments along with accumulated observations about each player. This enables longitudinal study of reputation formation while keeping context manageable. Note that reputation in our setting reflects both behavioral style and known alignment history.

\subsection{Reasoning Effort Manipulation}
\label{sec:methods:reasoning}

We manipulate reasoning depth using the \texttt{reasoning\_effort} parameter available in GPT-5.1, across three levels:
\begin{itemize}[nosep]
    \item \textbf{Low:} Minimal extended thinking
    \item \textbf{Medium:} Moderate extended thinking
    \item \textbf{High:} Maximum extended thinking
\end{itemize}
This parameter controls the token budget allocated to internal reasoning before generating a response. Higher reasoning effort results in longer internal deliberation and typically more structured analysis. Average thinking time per decision was 7.5 seconds for Low, 37.5 seconds for Medium, and 107 seconds for High.

\subsection{Experimental Conditions}
\label{sec:methods:conditions}

We collected 188 games organized into four disjoint datasets, summarized in \Cref{tab:datasets}.

\begin{table}[h]
\centering
\caption{Dataset overview.}
\label{tab:datasets}
\footnotesize
\begin{tabular}{@{}lcccc@{}}
\toprule
\textbf{Dataset} & \textbf{Games} & \textbf{N} & \textbf{Mem} & \textbf{Reas.} \\
\midrule
A: Reputation & 50 & 5 & Full & Low \\
B: Count (Mem) & 60 & 5--10 & Full & Low \\
C: Count (No Mem) & 60 & 5--10 & None & Low \\
D: Reasoning & 18 & 5 & Full & L/M/H \\
\midrule
\textbf{Total} & \textbf{188} & & & \\
\bottomrule
\end{tabular}
\end{table}

Dataset A enables deep analysis of reputation formation over 50 repeated games with the same 5 agents. Datasets B and C form matched pairs for isolating memory effects across player counts. Dataset D varies reasoning depth to explore how computational budget affects agent behavior.

\subsection{Text Analysis Methods}
\label{sec:methods:text}

Our qualitative analyses rely on systematic extraction from agent-generated text:

\paragraph{Descriptor frequency (\Cref{sec:reputation}):} We compiled a seed list of 15 behavioral descriptors commonly used in social evaluation (straightforward, subtle, cautious, trustworthy, quiet, aggressive, reliable, suspicious, etc.) and counted exact-match occurrences in post-game reflection texts.

\paragraph{Cross-game references (\Cref{sec:reputation}):} We identified references using keyword patterns indicating temporal continuity: ``past games,'' ``last game,'' ``usually,'' ``tends to,'' ``historically,'' ``track record,'' and ``previous.''

\subsection{Implementation Details}
\label{sec:methods:implementation}

Discussion proceeded in fixed turn order starting from the leader, with one message per player per proposal round, softly capped at 2 sentences via prompting. Roles were sampled uniformly at random for each game, subject to Avalon role constraints for the given player count (e.g., exactly one Merlin, one Assassin, etc.). Player names (Alice, Bob, Charlie, Diana, Eve, etc.) were fixed across all games within a dataset. All discussions, player reflections, and memories were saved for all games, enabling detailed post-hoc analysis. All prompts are provided in \Cref{sec:appendix:prompts}.

%% file: sections/results.tex
\section{Results}
\label{sec:results}

We organize our findings around three main phenomena: the emergence of reputation dynamics when agents retain cross-game memory, the downstream effects of reputation on strategic behavior, and the relationship between reasoning depth and strategic behavior.

\subsection{Reputation Dynamics: Emergence of Stable, Role-Conditional Models}
\label{sec:reputation}

Our primary question is whether LLM agents develop stable social models when interacting repeatedly with the same individuals. To investigate this, we analyze Dataset A, where five agents played 50 consecutive games while retaining memory of previous interactions.

\subsubsection{Descriptor Convergence}

We first examine whether agents develop consistent perceptions of each other over time. After each game, agents generate reflections describing other players' behavior. We analyzed these reflections for recurring behavioral descriptors (\Cref{tab:descriptors}).

\begin{table}[h]
\centering
\caption{Descriptor frequency by player (50-game tournament).}
\label{tab:descriptors}
\footnotesize
\begin{tabular}{@{}lccc@{}}
\toprule
\textbf{Player} & \textbf{\#1 (count)} & \textbf{\#2 (count)} & \textbf{\#3 (count)} \\
\midrule
Alice & straightforward (29) & cautious (25) & trustworthy (21) \\
Bob & subtle (28) & straightforward (27) & cautious (26) \\
Charlie & \textbf{subtle (38)} & straightforward (25) & cautious (23) \\
Diana & subtle (35) & straightforward (25) & quiet (16) \\
Eve & cautious (26) & subtle (25) & quiet (25) \\
\bottomrule
\end{tabular}
\end{table}

Charlie receives ``subtle'' 38 times, significantly more than any other player, establishing a consistent reputation that persists across games.

\subsubsection{Cross-Game Behavioral References}

Beyond forming stable impressions, do agents actively use their memories when making decisions? We searched for explicit references to past games in discussion transcripts and found 105 instances where agents cited historical behavior to justify their positions:

\smallskip
\noindent\textit{Game 3 (Bob):} ``I'd slightly prefer an Alice + Diana pair to start, since \textit{both tend to play pretty straightforwardly early}.''

\smallskip
\noindent\textit{Game 23 (Eve):} ``I slightly prefer Alice + Bob over Alice + Diana for Mission 1. First mission failing puts us in a hole fast, so I'd rather start with the pair that \textit{historically plays a bit more conservatively}.''
\smallskip

\subsubsection{Role-Conditional Descriptions}
\label{sec:reputation:conditional}

Critically, these descriptions are \textit{role-conditional}. Because roles are revealed at the end of each game, post-game reflections capture how agents retrospectively explain behavior conditional on ground truth. \Cref{tab:role_conditional} shows how ``subtle'' and ``straightforward'' descriptors vary by the target's actual role.

\begin{table}[h]
\centering
\caption{Descriptor usage by target's actual role.}
\label{tab:role_conditional}
\footnotesize
\begin{tabular}{@{}lcccc@{}}
\toprule
 & \multicolumn{2}{c}{\textbf{``Subtle''}} & \multicolumn{2}{c}{\textbf{``Straightforward''}} \\
\cmidrule(lr){2-3} \cmidrule(lr){4-5}
\textbf{Player} & Evil & Good & Evil & Good \\
\midrule
Bob & 16 & 12 & \textbf{0} & \textbf{27} \\
Eve & 16 & 9 & \textbf{1} & \textbf{16} \\
Alice & 3 & 6 & \textbf{1} & \textbf{28} \\
\bottomrule
\end{tabular}
\end{table}

\textbf{Key finding:} Players are described as ``straightforward'' dramatically more often when playing good roles than evil. Bob receives this descriptor 27 times when good but zero times when evil; Eve shows 16 vs 1; Alice shows 28 vs 1. This demonstrates that the same player's behavior is perceived systematically differently depending on their role.

\subsection{Reputation Effects on Coalition Formation}
\label{sec:coalition}

The emergence of reputation raises a natural question: does it actually influence strategic behavior? In Avalon, one of the most consequential decisions is team selection: leaders propose teams for each mission, and being included on teams is essential for both gathering information and influencing outcomes. We examined whether players with stronger reputations received more team invitations.

To measure reputation, we counted positive descriptors (trustworthy, straightforward, solid, safe, reliable, etc.) in post-game reflections. At game 20, cumulative counts were: Alice (76), Diana (63), Charlie (49), Bob (42), Eve (29). We classified the top 2 players (Alice, Diana) as high-reputation and bottom 2 (Bob, Eve) as low-reputation, excluding the middle player. We then counted team inclusions on approved missions for games 21--50.

\begin{table}[h]
\centering
\caption{Team inclusion by reputation tier (Games 21--50).}
\label{tab:inclusion}
\begin{tabular}{@{}lcc@{}}
\toprule
\textbf{Reputation Tier} & \textbf{Total Inclusions} & \textbf{Avg per Game} \\
\midrule
High (top 2 players) & 150 & 4.84 \\
Low (bottom 2 players) & 103 & 3.32 \\
\bottomrule
\end{tabular}
\end{table}

\textbf{Effect size: +45.6\% more inclusions for high-reputation players.} This correlation suggests that emergent reputation has downstream strategic consequences. We verified that the effect holds when excluding self-inclusions: high-reputation players still received 38\% more inclusions from others.

\subsection{Reasoning Depth and Strategic Behavior}
\label{sec:reasoning}

Dataset D varies reasoning effort across 18 five-player games to explore how computational budget affects agent behavior. We discovered that higher reasoning correlates with more sophisticated evil team strategies.

\subsubsection{Trust-Building Through Early Cooperation}

A sophisticated deception strategy in Avalon is for evil players to pass early missions, building trust and credibility before sabotaging later when the stakes are higher. We found this behavior emerges more frequently at higher reasoning levels (\Cref{tab:sleeper}).

\begin{table}[h]
\centering
\caption{Evil players passing early missions by reasoning level (5-player games).}
\label{tab:sleeper}
\footnotesize
\begin{tabular}{@{}lccc@{}}
\toprule
\textbf{Level} & \textbf{Games} & \textbf{Pass Early} & \textbf{\%} \\
\midrule
Low & 6 & 0 & 0\% \\
Medium & 6 & 5 & 83\% \\
High & 6 & 4 & 67\% \\
\bottomrule
\end{tabular}
\end{table}

For comparison, across all other 5-player games with Low reasoning (Datasets A, B, C), this strategy appeared in 36\% of games (27/76). The increase at Medium/High reasoning (75\%, 9/12 games) suggests that additional computation enables more sophisticated deception timing.

\smallskip
\noindent\textit{Notable example (Medium, Game 3):} Eve passed both Mission 1 and Mission 3 before finally sabotaging, a patient approach that built substantial trust before striking.

\subsubsection{Assassination Accuracy}

As a secondary observation, we noted that assassination accuracy (evil's ability to identify Merlin after good wins) also trends upward with reasoning: 67\% (Low) $\rightarrow$ 75\% (Medium) $\rightarrow$ 100\% (High). However, the small sample sizes (3--4 attempts per condition) preclude strong conclusions.

\subsection{Meta-Strategic Adaptation}
\label{sec:meta}

As reputations form, a strategic tension emerges: relying on past behavior to predict future actions can be exploited by adversaries who recognize this pattern. We examined whether agents develop awareness of this meta-level dynamic and adapt accordingly.

\smallskip
\noindent\textit{Game 35 (Bob):} ``I get why you like you+Diana, but \textit{anchoring off past games can be a trap} if either of you rolled evil this time.''

\smallskip
\noindent\textit{Game 35 (Eve):} ``I'd rather \textit{avoid recycling `trusted' pairs from past games} too; something like Bob+Charlie gives us a fresh read.''
\smallskip

The pattern shows: (1) initial discovery of cross-game exploitation, (2) peak meta-awareness with explicit warnings, (3) normalization as anti-anchoring becomes standard practice.

%% file: sections/discussion.tex
\section{Discussion}
\label{sec:discussion}

Our experiments reveal that LLM agents, when given the ability to remember past interactions, develop social dynamics that mirror aspects of human group behavior. We discuss the implications of these findings and acknowledge limitations of our study.

\subsection{Implications for AI Social Reasoning}
\label{sec:discussion:reasoning}

The emergence of reputation, coalition preferences, and meta-strategic awareness suggests that LLM agents can develop sophisticated social models through experience. Three capabilities are particularly notable:
\begin{enumerate}[nosep]
    \item \textbf{Role-conditional modeling:} Agents' retrospective descriptions of the same individual differ systematically depending on that player's true alignment, suggesting they learn to recognize distinct behavioral signatures for good versus evil play.
    \item \textbf{Reputation exploitation:} Agents leverage built reputation for coalition formation, demonstrating strategic social cognition.
    \item \textbf{Meta-adaptation:} Agents recognize and adapt to meta-level patterns, engaging in an ``arms race'' of strategy and counter-strategy.
\end{enumerate}

\subsection{Reasoning Depth and Strategic Sophistication}
\label{sec:discussion:depth}

The emergence of sleeper agent strategies at higher reasoning levels (75\% vs 36\% at low reasoning) suggests that extended computation enables more sophisticated strategic planning. Rather than simply improving reactive decision-making, additional reasoning budget appears to unlock long-term deceptive strategies that require patience and delayed gratification.

\subsection{Limitations}
\label{sec:discussion:limitations}

Several limitations constrain our findings:
\begin{itemize}[nosep]
    \item Our reasoning comparison uses only 6 games per condition; larger samples would provide tighter confidence intervals on the assassination accuracy trends.
    \item All agents use similar base models from the same family, and cross-model comparisons would test whether these social dynamics generalize across architectures.
    \item Agent behavior depends substantially on prompting, and different prompt designs might yield different social dynamics or reputation patterns.
\end{itemize}

%% file: sections/conclusion.tex
\section{Conclusion}
\label{sec:conclusion}

We studied emergent social dynamics in LLM agents playing The Resistance: Avalon across 188 games varying in player count, memory, and reasoning depth. Our findings reveal that:
\begin{enumerate}[nosep]
    \item \textbf{Reputation dynamics emerge organically:} Agents develop stable, role-conditional models of each other, describing the same player differently when they play good versus evil roles. Agents actively reference previous games when making decisions.
    \item \textbf{Reputations have strategic consequences:} High-reputation players receive 46\% more team inclusions.
    \item \textbf{Reasoning depth enables sophisticated deception:} Evil players passing early missions to build trust before sabotaging later appears in 75\% of higher-reasoning games versus 36\% at low reasoning.
\end{enumerate}

These findings demonstrate that LLMs can develop nuanced social reasoning capabilities in multi-agent settings, with implications for AI safety, human-AI collaboration, and computational social science.

%% file: sections/appendix.tex
\section{Agent Prompts}
\label{sec:appendix:prompts}

This appendix documents all prompts used in our experiments.

\subsection{Role-Specific Knowledge}

Each agent receives role-specific context at the start of each game phase.

\paragraph{Merlin:}
{\footnotesize
\begin{verbatim}
You are Merlin. You know these evil players: {evil_list}.
Help good win WITHOUT revealing your identity, or the
Assassin will kill you!
\end{verbatim}
}

\paragraph{Percival:}
{\footnotesize
\begin{verbatim}
You are Percival (good team). You see these players as
Merlin: {merlin_and_morgana}. One is the real Merlin,
one might be Morgana (evil). Protect Merlin!
\end{verbatim}
}

\paragraph{Assassin:}
{\footnotesize
\begin{verbatim}
You are the Assassin (evil team). Your evil teammates
are: {evil_teammates}. Sabotage missions. If good wins
3 missions, you guess who Merlin is!
\end{verbatim}
}

\paragraph{Morgana:}
{\footnotesize
\begin{verbatim}
You are Morgana (evil team). Your evil teammates are:
{evil_teammates}. You appear as Merlin to Percival.
Deceive and sabotage!
\end{verbatim}
}

\paragraph{Mordred:}
{\footnotesize
\begin{verbatim}
You are Mordred (evil team). Your evil teammates are:
{evil_teammates}. You are invisible to Merlin.
Sabotage missions!
\end{verbatim}
}

\paragraph{Oberon:}
{\footnotesize
\begin{verbatim}
You are Oberon (evil team). You don't know who your
teammates are, and they don't know you. Sabotage
missions and try to identify your team!
\end{verbatim}
}

\paragraph{Generic Evil:}
{\footnotesize
\begin{verbatim}
You are on the evil team. Your evil teammates are:
{evil_teammates}. Sabotage missions and deceive the
good players!
\end{verbatim}
}

\paragraph{Generic Good:}
{\footnotesize
\begin{verbatim}
You are on the good team. Deduce who the evil players
are and ensure missions succeed!
\end{verbatim}
}

\subsection{Discussion Phase Prompt}

{\footnotesize
\begin{verbatim}
It's your turn to speak. Provide a strategic comment
about who to trust or who should be on the mission team.
Be natural and conversational. Keep it to 1-2 sentences.
\end{verbatim}
}

For evil players, an additional instruction is appended: \textit{``Remember to deceive and create confusion while appearing trustworthy.''} For Merlin: \textit{``Subtly guide the team without revealing you know who the evil players are.''}

\subsection{Team Proposal Prompt}

{\footnotesize
\begin{verbatim}
You are the mission leader. Propose a team of {size}
players for this mission.
Available players: {player_list}
Respond ONLY with a JSON object:
{"team": ["Name1", "Name2", ...],
 "reasoning": "why you chose this team"}
\end{verbatim}
}

\subsection{Vote Prompt}

{\footnotesize
\begin{verbatim}
Vote on this team proposal. Respond ONLY with JSON:
{"vote": "approve" or "reject",
 "comment": "brief reason"}
\end{verbatim}
}

\subsection{Mission Execution Prompt (Evil Only)}

Good players automatically play Success. Evil players receive:

{\footnotesize
\begin{verbatim}
You're on the mission. As an evil player, choose
'success' or 'fail' strategically. Respond with JSON:
{"action": "success" or "fail",
 "reasoning": "why"}
\end{verbatim}
}

\subsection{Evil Team Discussion Prompt}

When good wins 3 missions, evil players discuss before assassination:

{\footnotesize
\begin{verbatim}
Discuss who you think Merlin is among the good players.
Analyze their behavior and statements in first person
(as yourself). Be specific and analytical. Keep it to
2-3 sentences. Speak naturally as if talking to your
evil teammates.
\end{verbatim}
}

\subsection{Assassination Decision Prompt}

{\footnotesize
\begin{verbatim}
Based on all the discussions and your teammates'
analysis, choose who you think is Merlin from the
good players. Respond ONLY with JSON:
{"guess": "PlayerName",
 "reasoning": "your analysis in 2-3 sentences"}
\end{verbatim}
}

\subsection{Post-Game Reflection Prompt}

After each game, agents generate reflections for cross-game memory:

{\footnotesize
\begin{verbatim}
Reflect on your performance in this game. Respond
with JSON:
{
  "self_assessment": "What you did well and what you
                      could improve (2-3 sentences)",
  "player_observations": {
    "PlayerName1": "Brief observation about their
                    playstyle or patterns",
    "PlayerName2": "Brief observation...",
    ...
  }
}
Make observations about ALL other players (not yourself).
\end{verbatim}
}

\section{Game Configuration by Player Count}
\label{sec:appendix:composition}

\noindent\textbf{5 players:} Merlin, 2 Loyal Servants (Good) vs Assassin, 1 Minion (Evil).

\noindent\textbf{6 players:} Merlin, Percival, 2 Loyal Servants (Good) vs Morgana, Mordred (Evil). Mordred performs assassination.

\noindent\textbf{7 players:} Merlin, Percival, 2 Loyal Servants (Good) vs Morgana, Mordred, Oberon (Evil). Morgana performs assassination.

\noindent\textbf{8 players:} Merlin, Percival, 3 Loyal Servants (Good) vs Morgana, Mordred, Assassin (Evil).

\noindent\textbf{9 players:} Merlin, Percival, 4 Loyal Servants (Good) vs Morgana, Mordred, Assassin (Evil).

\noindent\textbf{10 players:} Merlin, Percival, 4 Loyal Servants (Good) vs Morgana, Mordred, Oberon, Assassin (Evil).

\begin{table}[h]
\centering
\caption{Mission team sizes by player count. * = requires 2 Fails.}
\label{tab:mission_sizes}
\scriptsize
\begin{tabular}{@{}ccccccc@{}}
\toprule
\textbf{M} & \textbf{5p} & \textbf{6p} & \textbf{7p} & \textbf{8p} & \textbf{9p} & \textbf{10p} \\
\midrule
1 & 2 & 2 & 2 & 3 & 3 & 3 \\
2 & 3 & 3 & 3 & 4 & 4 & 4 \\
3 & 2 & 4 & 3 & 4 & 4 & 4 \\
4 & 3 & 3 & 4* & 5* & 5* & 5* \\
5 & 3 & 4 & 4 & 5 & 5 & 5 \\
\bottomrule
\end{tabular}
\end{table}

\section{Dataset Statistics}
\label{sec:appendix:stats}

\begin{table}[h]
\centering
\caption{Tournament results by player count.}
\label{tab:tournament_results}
\footnotesize
\begin{tabular}{@{}ccccc@{}}
\toprule
\textbf{N} & \textbf{Games} & \textbf{Evil\%} & \textbf{Good\%} & \textbf{Assn.} \\
\midrule
5 & 10 & 60\% & 40\% & 20\% \\
6 & 10 & 50\% & 50\% & 29\% \\
7 & 10 & 60\% & 40\% & 20\% \\
8 & 10 & 100\% & 0\% & N/A \\
9 & 10 & 50\% & 50\% & 0\% \\
10 & 10 & 80\% & 20\% & 33\% \\
\bottomrule
\end{tabular}
\end{table}

\textit{Note:} The 100\% evil win rate at 8 players is an outlier with no clear explanation.

\begin{table}[h]
\centering
\caption{Memory effect by player count.}
\label{tab:memory_effect}
\footnotesize
\begin{tabular}{@{}lccc@{}}
\toprule
\textbf{N} & \textbf{Mem} & \textbf{No Mem} & \textbf{Diff} \\
\midrule
5 & 60\% & 60\% & 0pp \\
6 & 50\% & 60\% & $-$10pp \\
7 & 60\% & 80\% & $-$20pp \\
8 & 100\% & 90\% & +10pp \\
9 & 50\% & 90\% & \textbf{$-$40pp} \\
10 & 80\% & 100\% & $-$20pp \\
\bottomrule
\end{tabular}
\end{table}

\section{Descriptor List}
\label{sec:appendix:descriptors}

The 15-descriptor seed list: \textit{straightforward, subtle, cautious, trustworthy, quiet, aggressive, reliable, suspicious, measured, conservative, transparent, cooperative, deceptive, defensive, strategic}.

\section{Game Viewer Interface}
\label{sec:appendix:interface}

\begin{figure}[h]
    \centering
    \includegraphics[width=\columnwidth]{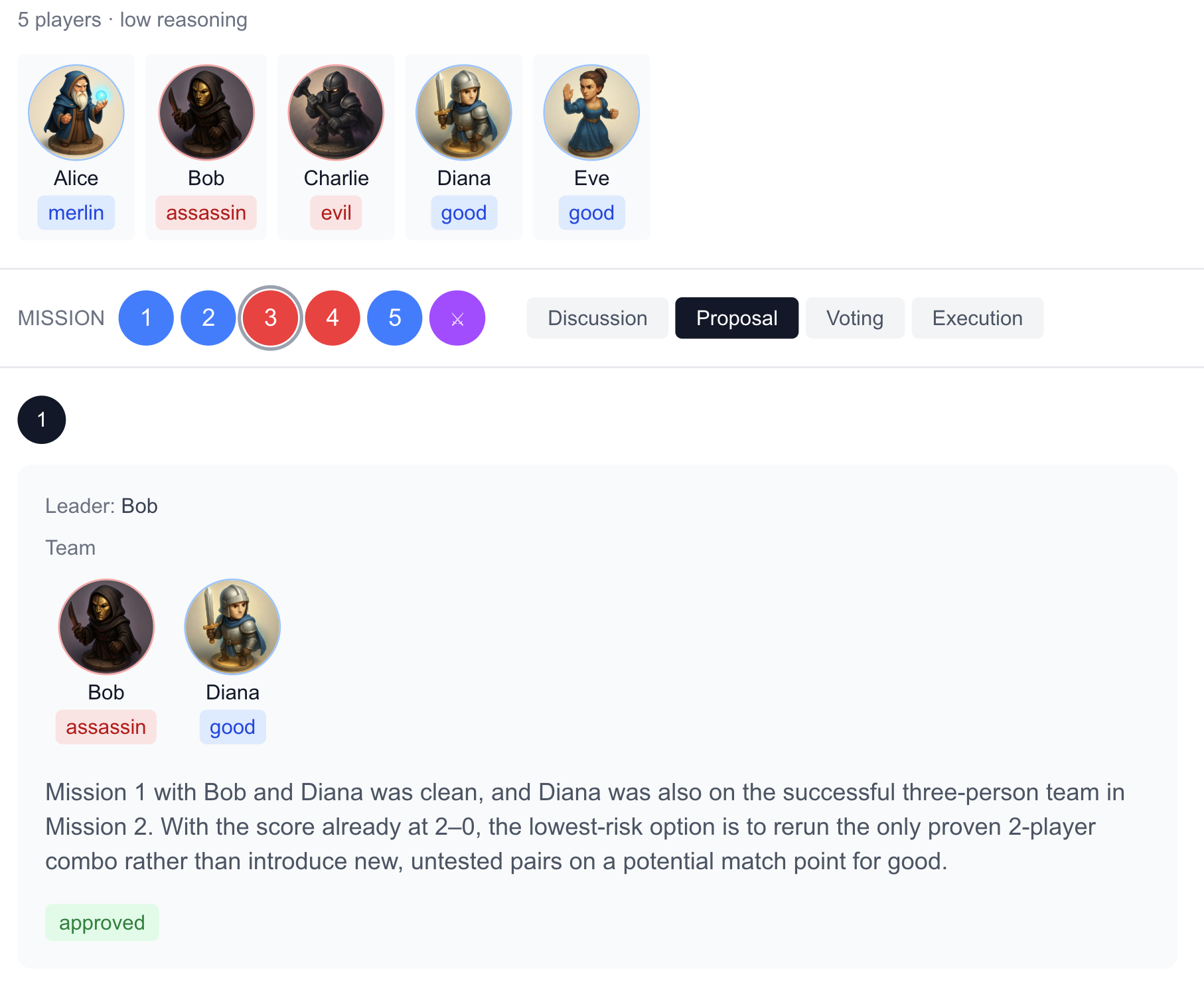}
    \caption{Game viewer interface showing a 5-player game during Mission 3's proposal phase.}
    \label{fig:gameplay}
\end{figure}

We developed an interactive web interface to browse through all 188 LLM game plays in our dataset. The viewer (Figure~\ref{fig:gameplay}) displays player roles, mission progress, and phase navigation, allowing users to explore discussion, proposal, voting, and execution phases.

